\documentclass[]{spie}  

\newcommand{\rhos}{\rho_\odot}
\newcommand{\rhop}{\rho_\oplus}
 
\usepackage{amsmath,amsfonts,amssymb}
\usepackage{graphicx}
\usepackage[colorlinks=true, allcolors=blue]{hyperref}

\title{RISTRETTO: coronagraph and AO designs enabling High Dispersion Coronagraphy at 2\:{\Large $\lambda/D$}}

\authorinfo{Further author information, send correspondence to N. Blind (nicolas.blind@unige.ch)}

\author[a]{N. Blind}
\author[a]{B. Chazelas}
\author[a,b]{J. K\"uhn}
\author[a]{E. Hocini}
\author[a]{C. Lovis}
\author[c]{M. Beaulieu}
\author[d]{T. Fusco}
\author[a]{L. Genolet}
\author[e,f,g]{O. Guyon}
\author[a]{J. Hagelberg}
\author[a]{I. Hughes}
\author[c]{P. Martinez}
\author[d]{J.-F. Sauvage}
\author[a]{R. Schnell}
\author[a]{M. Sordet}
\author[c]{A. Spang}

\affil[a]{Observatoire astronomique de l'Universite de Geneve, Sauverny, Switzerland}
\affil[b]{Space Sciences Institute, University of Bern, Bern, Switzerland}
\affil[c]{Observatoire de la C\^ote d’Azur, CNRS, Laboratoire Lagrange, Nice, France}
\affil[d]{DOTA,ONERA, Universit\'e Paris Saclay, Palaiseau, France}
\affil[e]{University of Arizona, Steward Observatory, Tucson, Arizona, United States}
\affil[f]{National Astronomical Observatory of Japan, Subaru Telescope, National Institutes of Natural Sciences, Hilo, HI96720, USA}
\affil[g]{Astrobiology Center, National Institutes of Natural Sciences, Osawa, Mitaka, Tokyo, JAPAN}

\begin{document}
\maketitle

\begin{abstract}
RISTRETTO is the evolution of the original idea of coupling the VLT instruments SPHERE and ESPRESSO \cite{lovis_2016a}, aiming at High Dispersion Coronagraphy. RISTRETTO is a visitor instrument that should enable the characterization of the atmospheres of nearby exoplanets in reflected light, by using the technique of high-contrast, high-resolution spectroscopy. Its goal is to observe Prox Cen b and other planets placed at about 35mas from their star, i.e. $2\lambda/D$ at $\lambda$=750nm. The instrument is composed of an extreme adaptive optics, a coronagraphic Integral Field Unit, and a diffraction-limited spectrograph (R=140.000, $\lambda =$620-840 nm). 

We present the status of our studies regarding the coronagraphic IFU and the XAO system. The first in particular is based on a modified version of the PIAA apodizer, allowing nulling on the first diffraction ring. Our proposed design has the potential to reach $\ge 50\%$ coupling and $\le 10^{-4}$ contrast at $2\lambda/D$ in median seeing conditions.
\end{abstract}


\section{INTRODUCTION}
\label{sec:intro}  

Direct detection of the light from extrasolar planets is a challenging objective. For young exoplanets on wide orbits this is achieved by high-contrast imaging and coronagraphy. For close-in exoplanets it is possible to take advantage of planetary transits and obtain a direct measurement of the IR flux of giant planets using secondary eclipses and phase curves. However, many known exoplanets including those orbiting the brightest and nearest stars remain out of reach of these techniques.

The RISTRETTO instrument will attempt reflected-light observations of spatially-resolved exoplanets for the first time. The proposed technique for RISTRETTO is High Dispersion Coronagraphy (HDC) spectroscopy, that can enable the $10^{-7}$ contrast level. To separate the planet from the star, adaptive optics on an 8m class telescope is used. In order to reach the required contrast, a coronagraph and an extreme AO stage are needed. Using a high-resolution spectrograph, the radial velocity shift between star and planet allows us to separate the stellar and planetary spectral lines, providing an additional factor $\sim$1000 in achievable contrast. Other experiments are being developed based on a similar approach. However they are targeting the thermal emission of young, massive planets at larger separations in the IR. Driven by the Proxima b science case, the RISTRETTO instrument will be targeting reflected light in the visible wavelength range. The short wavelengths allow us to spatially resolve Proxima b and other known very nearby exoplanets, placing them at about 2 $\lambda/D$ on a 8m-class telescope.

RISTRETTO will also use its high contrast, spectral and spatial resolution to study H-$\alpha$ emission lines (e.g. PDS70) or solar system objects.

\begin{figure}
    \centering
    \includegraphics[width=.8\textwidth]{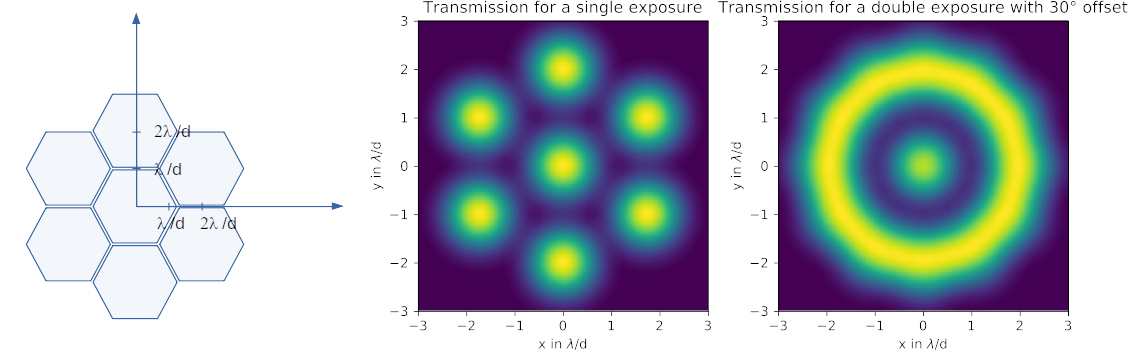}
    \caption{Left: schematic view of the RISTRETTO IFU. Midlle: transmission map due to single mode fiber coupling. Right: transmission map after $30^\circ$ rotation of the IFU.}
    \label{fig:ifu}
\end{figure}

\section{Requirements}
\label{sec:requirements}

In the case of a photon-noise limited measurement, the SNR of HDC technique is given by \cite{lovis_2016a}:
\begin{equation}
    \mathrm{SNR} \varpropto \sqrt{T} \dfrac{\rhop}{\sqrt{\rhos}},
\end{equation}
where $T$ is the total transmission, and $\rhop$ and $\rhos$ are respectively the planet and the stellar light coupled into the external fibers of the IFU. Note that in the rest of the paper, we will not consider contrast anymore, since it mixes the planet and star coupling, which are in practice optimized almost individually.

\noindent The requirements for RISTRETTO are the following:
\begin{enumerate}
    \item $\rhop \ge 50\%$ for stars as faint as PDS70 (Imag =11).
    \item $\rhos \le 10^{-4}$ for Proxima Cen
\end{enumerate}
, applying for median seeing conditions (Seeing = 0.83" at zenith distance of $30^\circ$, $L_0$ = 20m, Wind = 9.5m/s). Note that maximum elevation of Proxima Cen is 52deg, i.e. a corrected median seeing $\sim$0.90”.
Assuming a reasonable transmission $\sim$5\%, such requirements should allow a SNR = 5 on prox Cen in 40 nights.

In the original paper\cite{lovis_2016a}, ESPRESSO, with a resolution of R=220.000 was considered. The initial contrast requirement of $2\times10^{-4}$ (i.e. $\rhos = 10^{-4}$) was imposed by the limits in the current visible detectors RON and dark current, considering exposure times of 1h: going towards higher contrast would not improve the final SNR.
RISTRETTO is now designed with a goal R=140.000, meaning less pixels are required. Under the same assumptions made in \cite{lovis_2016a}, we show that for 1h integration time, the stellar signal will be equivalent to the detector RON if its coupling $\rhos \sim 3\times 10^{-5}$. Therefore, it would be meaningful to attempt pushing contrast to such extreme value to decrease integration times by up to 1.7 (which corresponds to days in the case of Prox Cen b).

\section{Integral Field Unit}
\label{sec:ifu}

The Integral Field Unit is meant to pave the VLT field of view, from 1 to 3 $\lambda/D$ distance. It is made of 7 hexagonal lenslets, 1 centered on the star, and 6 forming a ring at $2\lambda/D$ (Fig.~\ref{fig:ifu}). Each lenslet couples light into a single-mode fiber with cut-off in the visible (Throlabs SMF630HP). We do not consider multicore fibers for the moment, as we do not know if core-to-core contamination can be maintained sufficiently low. Because of the limited field-of-view of single-mode fibers, and the focus slicing operated by the lenslets, the transmission function of the IFU is limited, especially on the azimutal direction: if the planet is placed between exactly two lenslets, transmission drops to $\le$50\% from optimal. In order to optimize transmission it is therefore planned to rotate the IFU by 30$^\circ$ for half of the observing time, leading to an average $\sim$75\% geometrical transmission.

The IFU and fiber link are described in more details in \cite{kuhn_2022a}, with a particular focus on the current tests of our first prototypes.

\section{Coronagraph: a PIAA with nulling properties}
\label{sec:piaa}

\subsection{Principle}

\begin{figure}
    \centering
    \includegraphics[width=0.9\textwidth]{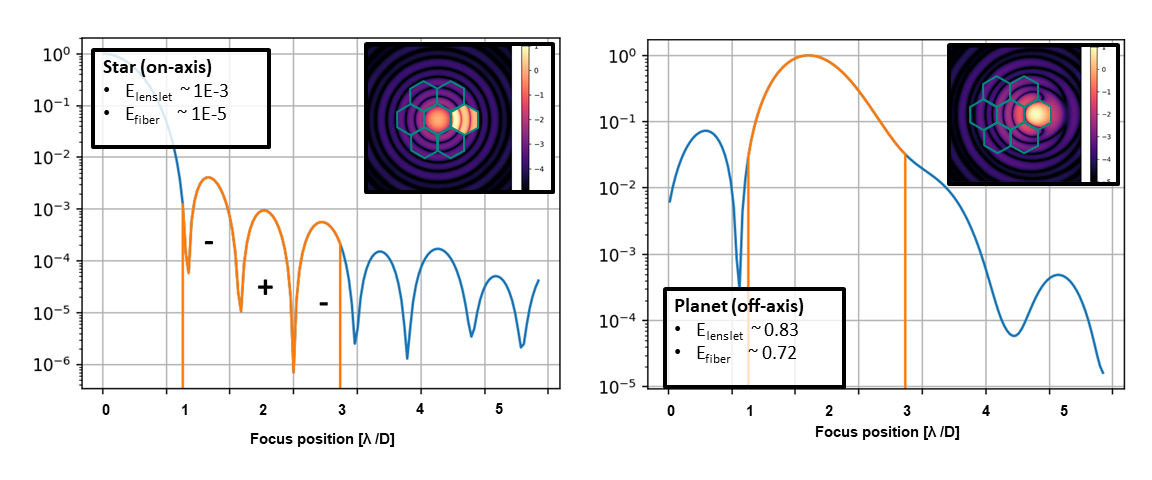}
    \caption{Example of post-PIAA PSF at $\lambda= 750$ nm. Left: on-axis (stellar) PSF: rings are attenuated by a factor $\ge 10$ compared to standard pupil and increase in frequency, providing some nulling level across a lenslet. Right: off-axis (planet at 2$\lambda/D$) PSF with moderate aberrations.}
    \label{fig:PIAA_principle}
\end{figure}

We developed for RISTRETTO the idea of a Phase Induced Amplitude Apodizer (PIAA) that would only partially apodize the pupil, and doing so provide some level of nulling of the star over the external single-mode fibers. This solution leads us to a win-win situation, where $\rhop$ is optimized and $\rhos$ is minimized.
\begin{itemize}
    \item For \textbf{stellar coupling}, we follow the idea of a nuller by partially apodizing the pupil, hence generating diffraction rings. Those rings are about 10 to 40 times fainter than in the unapodized case, and have higher spatial frequency, so that 2 (at $\lambda=840$nm) to 4  (at $\lambda=620$nm) are present in a lenslet. Because the electric field of thos rings is changing from one to the other, they eventually cancel each other over the SMF mode, hence creating nulls to the $10^{-5} - 10^{-6}$ levels (Fig.~\ref{fig:PIAA_principle}, left).
    \item \textbf{Planet coupling} (Fig.~\ref{fig:PIAA_principle}, right) is slightly improved by reducing energy contained in the diffraction rings, and which would be filtered out by the lenslet otherwise. For the current designs, we generally get about 80-85\% of the flux into the lenslet at 2$\lambda/D$. The PIAA off-axis aberrations are reasonable thanks to the moderate apodization.
    The electric field transmitted by the lenslet is still a rather good match to the SMF mode, so that the coupling from lenslet to fiber is around 90\%. This leads to $\rhop^0 \sim 70-75\%$ for the best solutions, slightly higher than with a VLT pupil with similar IFU. Considering AO residuals, we can also deduce that $\rhop = \rhop^0 \times \mathcal{S}$, where $\mathcal{S}$ is the Strehl ratio.
\end{itemize}
This makes our PIAA design one of the highest transmission coronagraph with inner working angle of 1.5 to 2.5$\lambda/D$ and band-pass of $\sim 30\%$. The higher transmission of the PIAA is a fundamental advantage once on sky, since solutions with lower transmission lead to more stringent constraints on the Strehl, i.e. the actuator count and/or seeing conditions.
\subsection{Optimization}

\begin{figure}
    \centering
    \includegraphics[width=1.\textwidth]{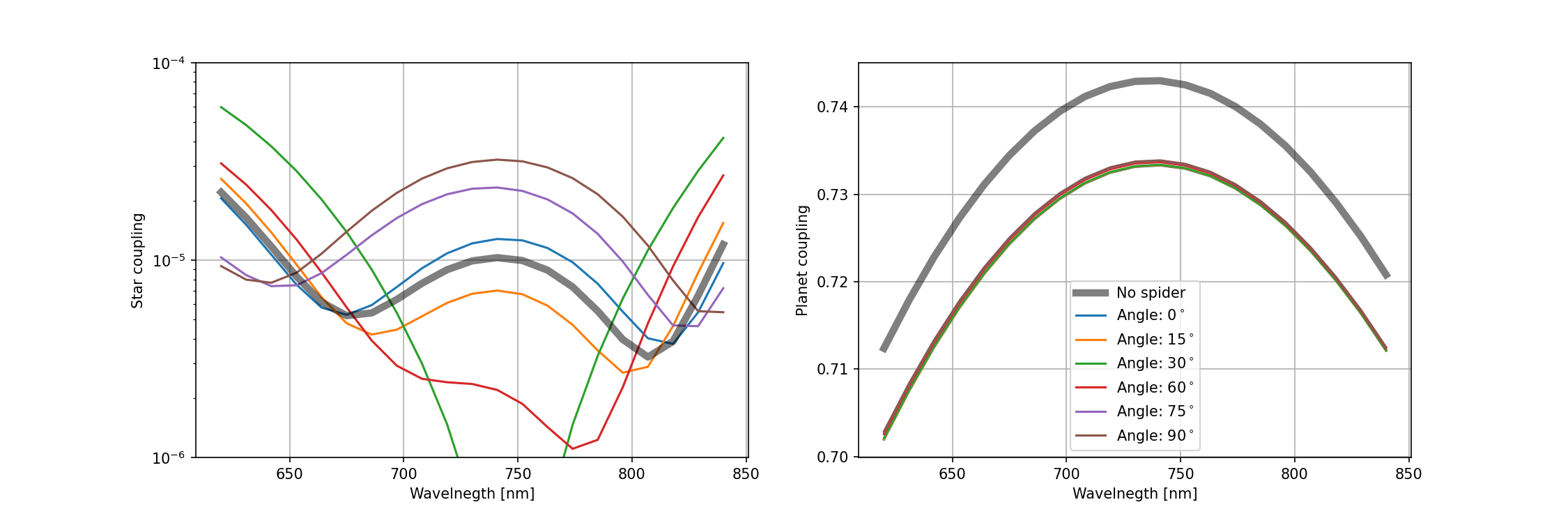}
    \caption{Stellar and planet coupling as a function of wavelength for a CaF2 PIAA. We show the case of VLT pupil (16.2\% obstruction) without and with spiders, under different orientations with respect to the lenslet.}
    \label{fig:piaa_perf_lbd}
\end{figure}

The optimization is performed over 4 parameters:
\begin{enumerate}
    \item Pupil illumination FWHM after PIAA -- While a small FWHM value (i.e. a more aggressive PIAA) will attenuate the rings and concentrate more energy in the PSF core, it will also lead to higher off-axis aberration of the PIAA, even at 2$\lambda/D$. 
    \item The filling factor of the central obstruction by the PIAA -- Filling the secondary obstruction plays a quite crucial role, since it also participates to determine the intensity of the diffraction rings. Those are proportionally less intense than the FWHM is small, since more energy is again concentrated in the center of the pupil. In practice, reducing the VLT secondary obstruction to 20\% of its initial value (i.e. $\sim$3\% of telescope diameter) appears necessary to optimize star rejection. Due to diffraction between PIAA optics, no benefit was observed for smaller values.
    \item Lenslet size -- The lenslet constitutes a spatial filter and is used to optimize primarily the null level. We only considered hexagonal lenslet on the external ring for the moment, lthough new 3D printing techniques may allow us some flexibility.
    \item The lenslet focal lens -- It is a weak free parameter, although it allows some fine tuning of the null level. 
\end{enumerate}
Of the 4 free parameters, the FWHM and the lenslet size are the main ones.  We started our studies by using an analytic solution, considering a gaussian apodization function. We will also consider other functions in the future.

PIAA simulations are using the Hcipy Python library\cite{por_2018a}, from telescope to fiber coupling. We started our studies with a toy-model based on an analytical solution, for which a post-PIAA phase remapping function is generated. This is equivalent to consider only Fraunhoffer propagation. We now have developed a Fresnel model, including the case of refractive lenses. The toy model generates more optimistic results, with nulls down to $10^{-6}$.

Our best solution at the moment reaches $\rhop \sim 72\%$ and $\rhos \sim 10^{-5}$ over the required 30\% band-pass and considering a CaF2 lens solution (Fig.~\ref{fig:piaa_perf_lbd}). A mirror solution provides marginally better performance. Note that we always show performance for a single lenslet, and not for the average of 6. Fig.~\ref{fig:piaa_perf_lbd} presents results with and without spiders, and for different orientations with respect to the lenslet in the latter case. We can observe the strong orientation and wavelength dependency of $\rhos$, i.e. of SNR. Such effects are largely smoothed by the RON limit and the AO residuals in practice, with SNR variations $\sim 2$ over band-pass and lenslets for best AO simulations (Sect.~\ref{sec:xao}). A first prototype should be made based on that solution.

As explained in Sect.~\ref{sec:requirements}, $\rhos \le 3\cdot 10^{-5}$ has limited interest for performance, since RON will be dominating signal. Down to $\rhos\sim10^{-6}$, such nulls however provide slightly higher tolerance of the system. Such contrasts have only been achieved with the toy-model nevertheless.

\subsection{Tolerances}

After finding proper solution, we analyzed tolerances with respect to manufacturing, alignment or wavefront errors. Tolerance are given for individual terms with a limit at $\rhos < 3\cdot10^{-5}$ to keep margin for others sources.

\subsubsection{Tip-tilt sensitivity}

\begin{figure}
    \centering
    \includegraphics[width=1.\textwidth]{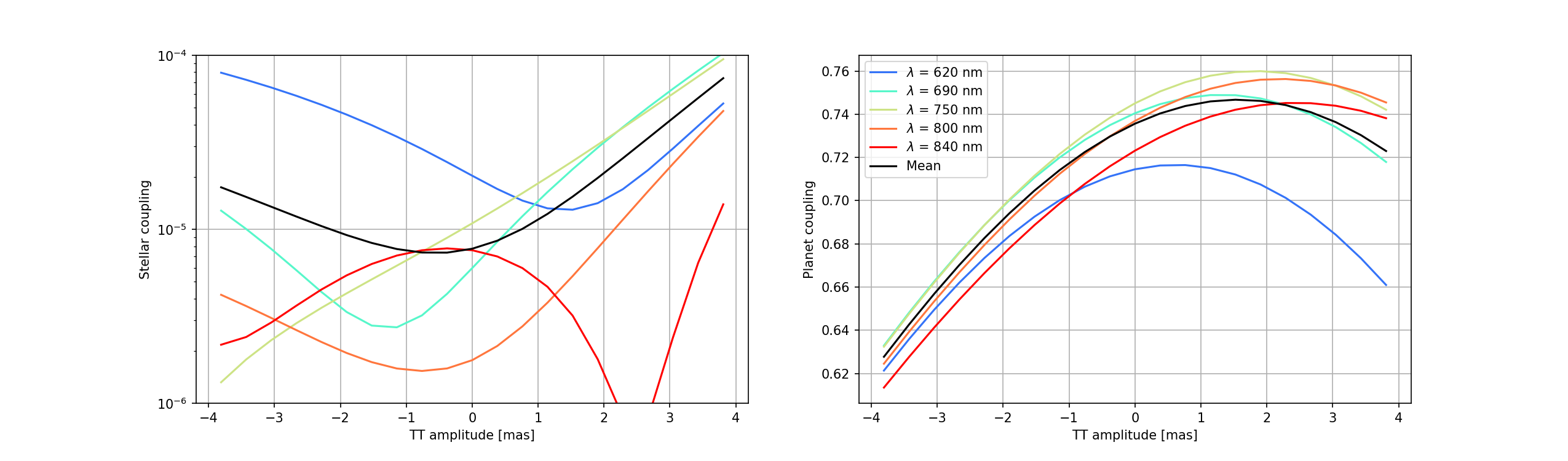}
    \caption{Tip-tilt sensitivity for a spider-free VLT aperture. Negative tilt pushes the star further away from the lenslet.}
    \label{fig:TT_sensitivity}
\end{figure}

Tip-tilt tolerance range extends from -4 to +2 mas (i.e. -0.2 to +0.1 $\lambda/D$; see Fig.~\ref{fig:TT_sensitivity}, left). The behavior is quite chromatic, although this should be smoothed in practice by AO residuals. Planet coupling is optimized for a slightly off-axis position, which suggests the lenslet shape  could be slightly optimized. Prox Cen diameter being $\sim 1$ mas, it will not impact the nuller performance.

\subsubsection{Wavefront errors}

Following \cite{por_2018b}, we estimated the sensitivity of our coronagraph to optical aberrations. We injected small wavefront errors (e.g. DM influence functions) and compute the change in the star electric field, the dominant term regarding SNR degradation. This allows to build an 'EF interaction matrix' and to extract the eigen aberrations of the coronagraph. This method is accurate for estimating null degradation only around a true null, but less otherwise (e.g. the modes ordering is false, or 'null' can be deepen).

We performed the analysis on 1 and 6 fibers at the same time, which naturally leads to different eigen modes, and potentially allows for a different optimization of the AO system, e.g. if we know the planet position. 
We are mostly dominated by 6 modes, with geometry matching the 7-fiber IFU one. The results are also similar (at least visually) to those of \cite{por_2018b}, which is considering as well a nuller and similar IFU geometry. Applied to another type of nuller (a binary ring apodizer in pupil plane\cite{kuhn_2022a}), we observed again similar results. Such aberration are well contained within 3 pupil cycles.

\begin{figure}
    \centering
    \includegraphics[width=\textwidth]{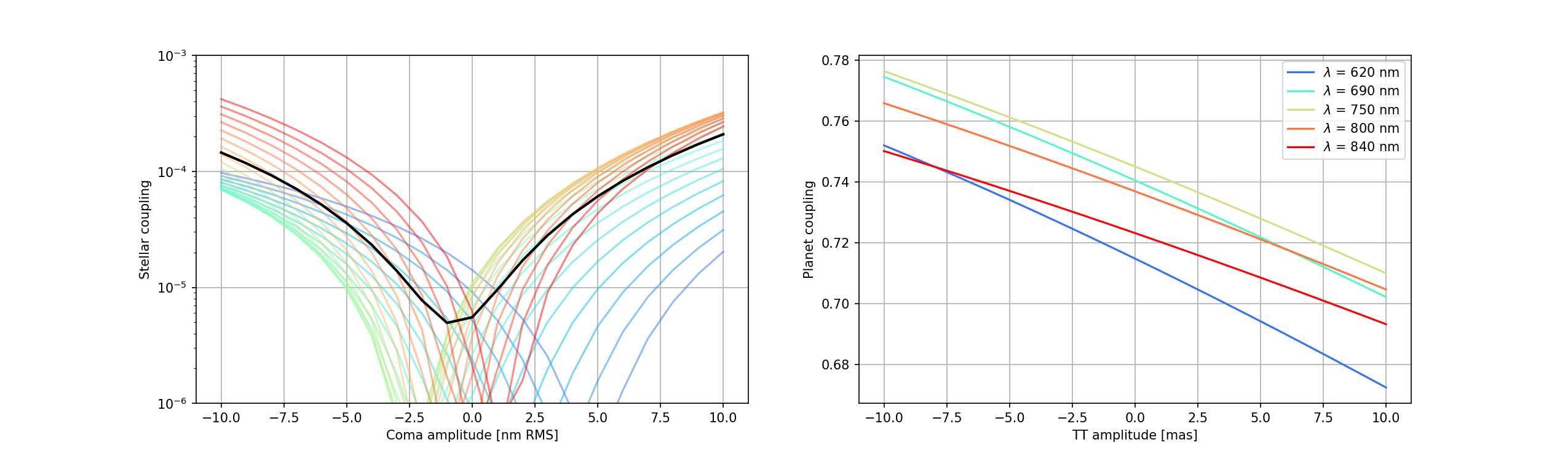}
    \caption{Sensitivity to coma aberration for a spider-free VLT aperture. Coloring on the left plot follows the logic of the right plot, with finer wavelength sampling. Black curve on the left is the average over the band.}
    \label{fig:coma_sensitivity}
\end{figure}

We more generally perform the analysis regarding coma since it is an aberration easily created by optical misalignment, and which proved generally the most sensitive classical aberration. Fig.~\ref{fig:coma_sensitivity} shows an average tolerance of about $\pm 5nm$, with again very chromatic behavior: some parts of the spectrum become very quickly blind due to stellar leakage. We remind that the analysis is performed on a single lenslet: for the one placed on the opposite side of the star, the behavior is reversed, which could make the search of Prox Cen b problematic. The red part of the spectrum presents less tolerance, as seen from the width of the curve, probably because of the lower ring count ($\sim 2$) across the lenslet. We also note the linear evolution of coupling for the planet, suggesting we are compensating the PIAA off-axis aberration, at the expense of contrast.

Computation of coma tolerance for many different PIAA configurations (with the toy model only) does not suggest that a fundamental improvement is possible by a wise optimization, at least not with the considered free parameters. It however tends to show that aiming at nulls down to $\rhos \sim 10^{-6}$ improves slightly tolerances.

\subsubsection{Dead actuators}

We considered a fixed actuator creating a few $\mu m$ wavefront bump in the DM. We hide it  behind a spider, and add a mask covering its direct neighbours, which should extend the bump via mechanical coupling. The additional diffraction pattern the mask generates significantly distorts the diffraction rings in the external lenslets, severely affecting null performance. 2 dead actuators in the pupil is not an acceptable situation. We have not explored yet an optimization if such situation should appear, but dead, fixed actuator(s) represents a risk.

\subsubsection{Comparison to other coronagraphs}

We performed a similar tolerance study for other coronagraphs, namely: a perfect coronagraph, a vortex and a specially optimized ring apodizer. All were slightly optimized in order to provide null depths of the order of $10^{-5}$, to the detriment of transmission if necessary. They all led to similarly tight tolerances, suggesting this is a property of the contrast requirements, and are not a property of the PIAA-nuller alone. PIAA actually presents higher tolerances: wavefront errors generate 'leaky electric field' (deviation from the ideal one) that still presents the property of fainter and higher frequency rings, and therefore also a higher potential to partially cancel out over the SMF.

\section{XAO design and expected performance}
\label{sec:xao}

\subsection{Requirements and error budget}

Following the PIAA analysis, we can deduce the following requirements on the wavefront error budget:
\begin{itemize}
    \item Strehl($\lambda$=750nm) $\ge$ 70\% or WFE $\le$ 70 nm2;
    \item Low order WFE, within 3 cycles, $\le$ 10nm RMS;
\end{itemize}
Note that for cases like PDS70, Strehl should be maintained up to Imag$\sim 11$.

If we break down the AO error budget in the usual terms, we can get a first idea of the minimum system we are aiming at. We consider a simple integrator controller, and Paranal median seeing conditions: 0.75" at zenith and wind speed of 10m/s. This corresponds to an equivalent seeing of 0.85" for Prox Cen b at its maximum elevation of 52$^\circ$.
\begin{itemize}
    \item \textbf{Fitting error} requires at least 32 actuators across the pupil to reach the Strehl requirement.
    \item \textbf{Lag error} requires at least a correction frequency of 1.5kHz to reach the low order WFE requirement. Integrated over all frequencies its contribution amounts to 17nm RMS, and is therefore a small contributor to Strehl.
    \item \textbf{WFS noise} influences mostly low order WFE in our case. Pyramid WFS is necessary and modulation amplitude plays an important role. Tab.~\ref{tab:flux_WFS} presents the minimum flux level required to reach this requirement, which do not necessarily ensure Strehl ratio is reached. Minimum flux ns the unmodulated case would be about 4000 ph/frame, while standard 3$\lambda/D$ requires 10 times more. 
    \item \textbf{Aliasing} plays a negligible role in our case.
\end{itemize}

\noindent In the case of unmodulated pyramid, servo-lag is largely dominating the error budget at low spatial frequencies $\le 3\lambda/D$. For modulation of 2 to 3 lbd/D, WFS noise is on par with the lag term at low spatial frequencies.

\begin{table}[b]
    \centering
    \caption{Minimum flux level on a 4-sided pyramid WFS to ensure WFE $\le$ 10nm RMS within 3 cycles, for different modulation amplitudes. Readout noise is negligible.}
    \label{tab:flux_WFS}
    \begin{tabular}{ll|ccccc}
        Modulation radius & [$\lambda/D$] & 0 & 1 & 2 & 3 & 5\\
        \hline
        Minimum flux & [e-/fr] & 4$\cdot 10^3$ & 8$\cdot 10^3$ & 20$\cdot 10^3$ & 40$\cdot 10^3$ & 170$\cdot 10^3$ \\ \\
    \end{tabular}
\end{table}

\subsection{Bandpass and chromatic errors}

\begin{figure}
    \centering
    \begin{minipage}{0.48\textwidth}
    \centering
    \includegraphics[width=1.\textwidth]{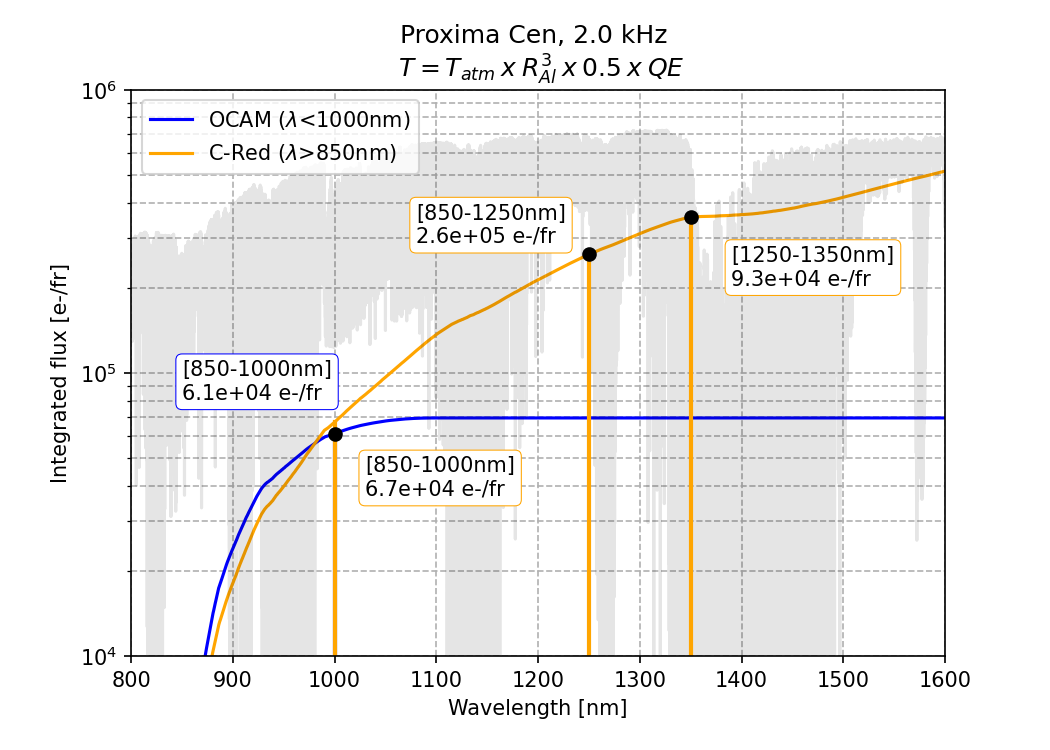}
    \caption{Photon budget on Prox Cen for 2kHz system.}
    \label{fig:wfs_photon}
    \end{minipage}
    \begin{minipage}{0.48\textwidth}
    \centering
    \includegraphics[width=1.00\textwidth]{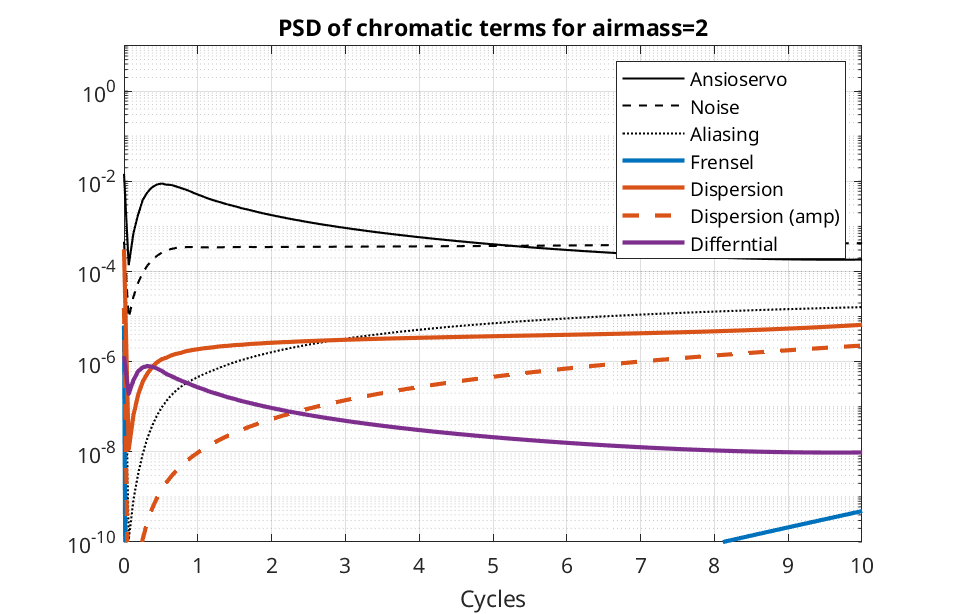}
    \caption{AO PSDs profiles for the different term in the AO error budget for a 50x50, 4kHz system under good seeing and airmass=2. We consider the worst case with $\lambda_{SCI}$=620nm and $\lambda_{WFS}$=1400nm.}
    \label{fig:psd}
    \end{minipage}
\end{figure}

The optimal bandpass and zero-point for the WFS are determined by using a PHOENIX model of Prox Cen (\texttt{lte02300-5.00-0.0.PHOENIX-ACES-AGSS-COND-2011-HiRes}). For transmission factors, we consider:
\begin{itemize}
    \item TAPAS atmospheric transmission model;
    \item 3 Al mirrors for the VLT;
    \item 50\% optical transmission from telescope focus to WFS for $\lambda\ge$850nm (0 otherwise);
    \item WFS camera QE, considering First Light Imaging cameras OCAM2K or C-Red 1;
\end{itemize}
Up to 1000nm, flux available is about $6\cdot10^4$ ph/fr (Fig.~\ref{fig:wfs_photon}), just slightly above what is required for a modulated pyramid WFS regarding low orders. This also does not allow for maintaining Strehl performance for I=11 (PDS70 case). We therefore need to extend the sensitivity band up to 1250nm to quadruple the flux to $26\cdot10^5$ ph/fr. Extending to redder parts becomes less and less interesting, while significantly adding complexity and risk in terms of chromaticity. Depending on cases, we may consider different filters for the WFS.

Fig.~\ref{fig:psd} presents the AO PSDs for Prox Cen with a highly performing system (50x50 actuators, 4kHz, 2 frame delay, unmodulated pyramid) under a good seeing, showing the balance between WFS noise and servo-lag error, the latter still dominating near the PSF core. We also considered unfavorable chromatic terms, between $\lambda_{SCI}=620$nm and $\lambda_{WFS}=1400$nm, and with an airmass=2, showing they should be negligible in our case.

\subsection{Performance}

We performed AO Fourier simulations with the Matlab OOMAO package\cite{conan_2014a}, producing a set of 100 independent phase screens for each. Those screens are injected into our PIAA+IFU model to estimate real performance of the system. A large grid was studied, with varying number of actuators, loop speed, pyramid modulation amplitude, or different seeing conditions. We consider a banpass up to 1250nm (i.e. $\sim 5\times10^5$ ph/ms), and chromatic effects are for the moment not included.

\begin{figure}
    \centering
    \includegraphics[width=\textwidth]{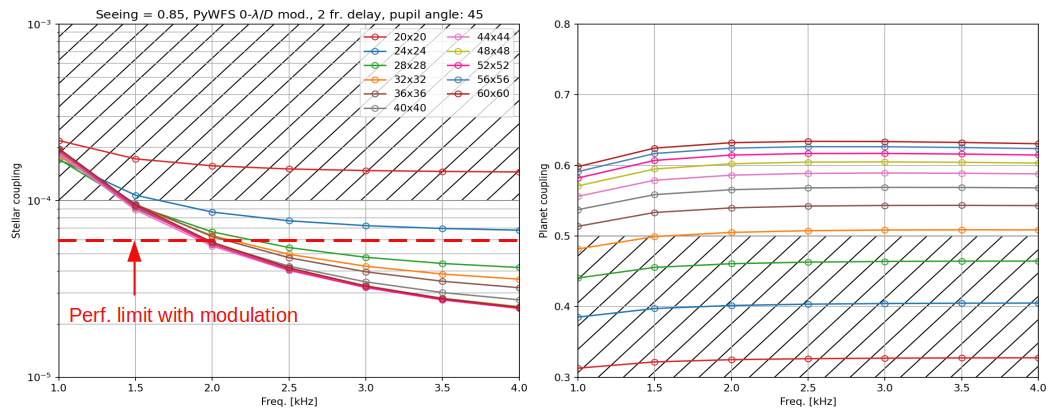}
    \caption{End-to-end RISTRETTO performance on Prox Cen as a function of loop speed. We consider a VLT pupil with spiders oriented to show an average behavior. On the left, the red dotted line shows the performance limit with modulation of 3$\lambda/D$.}
    \label{fig:AO_perf}
\end{figure}

With the PIAA design presented previously, the minimum system design appears as 32x32 actuators, 2kHz loop speed, and still considering a simple integrator (Fig.~\ref{fig:AO_perf}, for unmodulated pyramid), in good agreement with previous estimates. Strehl is fully constrained by number of actuators, while we can push contrast performance by increasing loop speed and decreasing the servo-lag error in the case of an unmodulated pyramid. As soon as we modulate the pyramid the loss of sensitivity on low orders limits performance for $f \ge$ 2kHz.

Contrast will likely be very difficult to achieve and maintain in practice, so that a higher order system (i.e. a Strehl focused) is favored. Strehl gain with higher actuator count saturates above 50x50 actuators, but guarantees 'requirement' performance for lower seeing conditions, up to seeing 1.2", i.e. for more than 80\% of the nights at Paranal.

\begin{figure}
    \centering
    \includegraphics[width=0.6\textwidth]{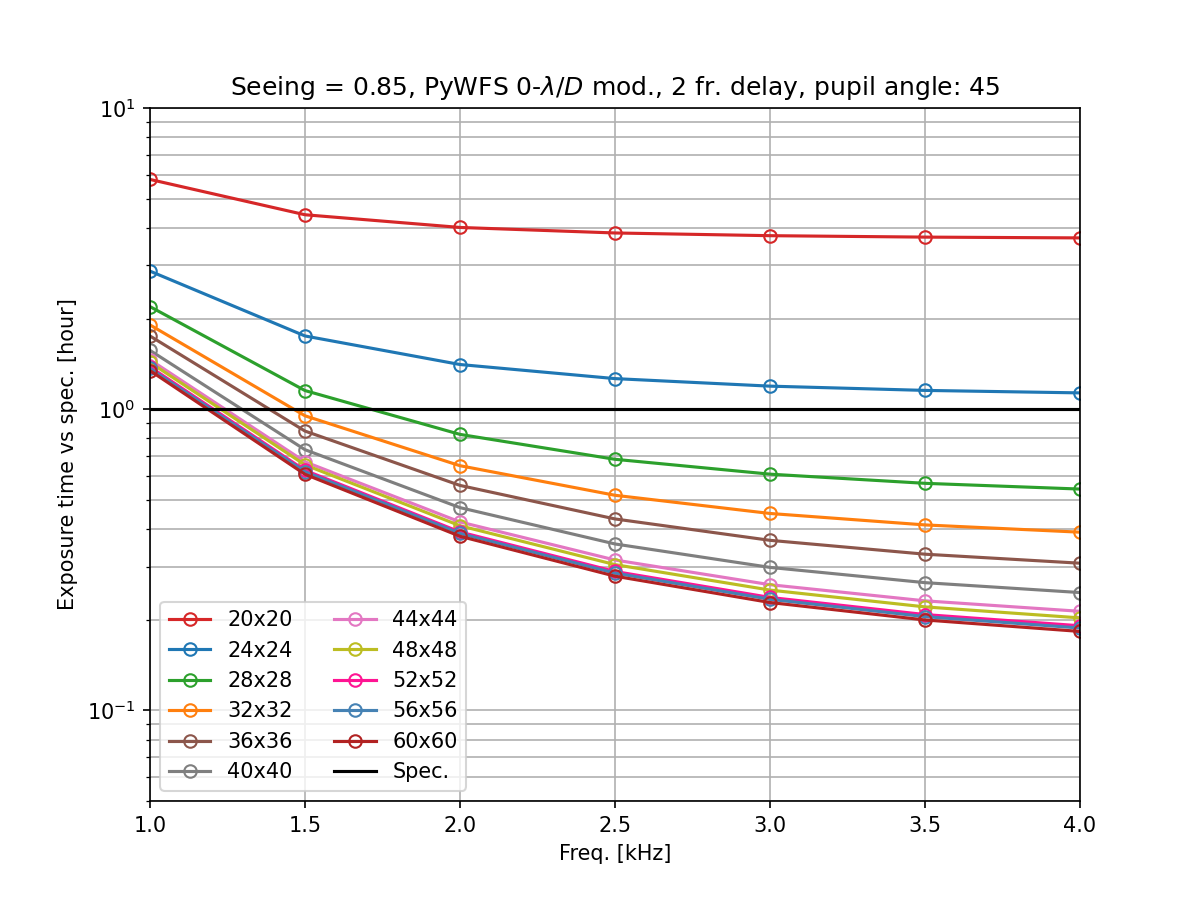}
    \caption{Potential gain in exposure time with respect to the requirement for the different AO configurations.}
    \label{fig:exp_time}
\end{figure}

Our baseline XAO system is currently 40x40 actuators and 2kHz pyramid WFS (modulated or not).
Fig.~\ref{fig:exp_time} finally shows the exposure time relative to the requirement for the previous simulations: this baseline design can potentially lead to up to 40\% shorter exposure times, i.e. saving 10 VLT nights or more, if achieved.

\section{Conclusion}
\label{sec:ccl}

We converged to an XAO and coronagraphic IFU design that enables our science goals, in particular with goal to detect Prox Cen b.  Our baseline XAO and PIAA design performance promise flexible scheduling and/or saving many VLT nights. Some optimization are probably still possible on the PIAA, e.g. considering optimized lenslet shape, always keeping in mind the technological limits. We are working on tolerancing a first PIAA prototype to assess them.

Following those results, we also started thinking to a preliminary Front-End concept. A few other challenges are already identified, in particular:
\begin{itemize}
    \item Understanding of the fundamental limitations of the pyramid WFS class, modulated or not.
    \item The ADC must reach about 1-mas correction level. Based on models, a design was proposed [private comm.]. However, at such level we start questioning the accuracy of models and what parameters to use for setting. Closed-loop ADC will be studied.
    \item Polarisation effects must be studied. They could be an important limiting factor for any attempt at high contrast in the first diffraction rings.
    \item Measurement and control of low order aberrations to the nm level over time scale of 1 hour, if not longer.
\end{itemize}

We are now moving towards prototyping and ordering of some identified key components.

\section*{Acknowledgement}
This work has been carried out within the framework of the National Centre of Competence in Research PlanetS supported by the Swiss National Science Foundation under grants 51NF40\_182901 and 51NF40\_205606. The RISTRETTO project was partially funded through SNSF FLARE programme for large infrastructures under grants 20FL21\_173604 and 20FL20\_186177. The authors acknowledge the financial support of the SNSF.

\bibliography{references} 
\bibliographystyle{spiebib} 

\end{document}